\documentclass[9pt,article,twoside]{rmaa-rho-class/rmaa-rho}
\RMxAAtemplatetype{\RMxAA} 

\makeatletter
\providecommand{\@doi}{}
\makeatother

\vol{-}
\pages{-}
\thisyear{2026}

\title{Wind accretion onto planets orbiting an evolving  Solar-like star and their detectability}

\author[1,2]{P.~Padilla-López\orcidlink{0009-0003-2981-2678}}
\author[2]{R.~F.~Maldonado\orcidlink{0000-0002-2236-7554}}
\author[1,2]{J.~A.~Toalá\orcidlink{0000-0002-5406-0813}}
\author[3]{E.~Tejeda\orcidlink{0000-0001-9936-6165}}
\author[4]{J.~B.~Rodríguez-González\orcidlink{0000-0002-0616-8336}}

\affil[1]{Facultad de Ciencias de la Tierra y el Espacio, Universidad Aut\'{o}noma de Sinaloa, Josefa Ortiz de Dom\'{i}nguez S/N, Culiac\'{a}n 80040, Sinaloa, Mexico}
\affil[2]{Instituto de Radioastronomía y Astrofísica, Universidad Nacional Autónoma de México, 58089 Morelia, Michoacán, Mexico}
\affil[3]{SECIHTI - Instituto de Física y Matemáticas, Universidad Michoacana de San Nicolás de Hidalgo, Ciudad Universitaria, 58040 Morelia, Michoacán, Mexico}
\affil[4]{Instituto de Astrofísica de Andalucía, CSIC, Glorieta de la Astronomía S/N, Granada 18008, Spain}

\leadauthor{P.~Padilla-López et al.}
\smalltitle{LaTex Macro RMxAA}

\corres{P.~Padilla-López}
\email{paulinapadilla.facite@uas.edu.mx}

\received{-}
\accepted{-}

\license{Texto de la licencia aquí}

\setbool{rho-abstract}{true} 
\setbool{rho-resumen}{true} 

\begin{abstract}
\noindent As stars evolve, they undergo significant changes in their physical properties, which can have a profound impact on the planets orbiting them. In particular, the mass lost through stellar wind may be partially accreted by orbiting planets. We present the results of 18 simulations of one-planet systems with planetary masses of 0.5, 1, 2.5, 5, 10, and 13~$\mathrm{M}_\mathrm{J}$, each at initial orbital distances of 5, 10, and 20~AU, orbiting a 2~M$_\odot$ star through its red giant branch and thermally pulsating asymptotic giant branch phases. Our results show that planets with smaller orbits and higher masses accrete more stellar wind material than their wider-orbit and lower-mass counterparts, although the total mass accreted across all simulations remains small compared to their initial planetary mass. Even for the most massive planet, 13 $\mathrm{M}_\mathrm{J}$ at 5 AU, the total mass accreted was $\sim0.56$\% of the planet's initial mass; nevertheless, we find that the accretion luminosities of the simulated planets, with the exception of one planet,  exceed their expected equilibrium luminosities, suggesting that such emission could be potentially detected. This result is key for the detection of planets around AGB stars, which have no confirmed detections as of yet. We also estimated the accretion and luminosities of two detected two-planet systems over a few orbits, obtaining results consistent with the one-planet simulated systems. Additional tests without wind accretion and with stellar wind drag force showed that, while both have a negligible effect on the orbital evolution, wind accretion remains relevant for the planetary luminosity. 
\end{abstract}


\begin{resumen}
A medida que las estrellas evolucionan, experimentan cambios significativos en sus propiedades físicas, lo que puede afectar de manera importante a los planetas que las orbitan. En particular, la masa perdida a través del viento estelar puede ser parcialmente acretada por los planetas. Presentamos los resultados de 18 simulaciones de sistemas con un solo planeta, con masas de 0.5, 1, 2.5, 5, 10 y 13~$\mathrm{M}_\mathrm{J}$, a distancias orbitales iniciales de 5, 10 y 20~UA, orbitando una estrella de 2~M$_\odot$ durante las fases de la rama gigante roja y la rama asintótica de las gigantes con pulsos térmicos. Nuestros resultados muestran que los planetas más masivos y en órbitas más cercanas acretan más material del viento estelar, aunque la masa total acretada en todas las simulaciones es pequeña en comparación con la masa inicial planetaria. Incluso para el planeta más masivo, de 13~$\mathrm{M}_\mathrm{J}$ a 5~UA, la masa total acretada fue de $\sim0.56$\%; sin embargo, las luminosidades por acreción de los planetas simulados, con la excepción de uno, superan sus luminosidades de equilibrio, lo que sugiere que esta emisión podría ser detectable. Este resultado es clave para la detección de planetas alrededor de estrellas AGB, las cuales aún no tienen detecciones confirmadas. Además, estimamos las tasas de acreción y las luminosidades de dos sistemas observados con dos planetas durante unas pocas órbitas, obteniendo resultados consistentes con los sistemas de un solo planeta. Pruebas adicionales sin acreción de viento y considerando la fuerza de arrastre del viento estelar indican que, aunque ambos efectos tienen un impacto despreciable en la evolución orbital, la acreción de viento sigue siendo relevante para la luminosidad planetaria.
\end{resumen}

\begin{document}

\maketitle
\pagestyle{fancy}\thispagestyle{firststyle}

\section{INTRODUCTION}
\label{intro}

While the first exoplanet was discovered orbiting a pulsar \citep{1992Natur.355..145W}, detections of planets around evolved giant stars have been consistently reported since the early 2000s \citep{2002Frink}. To date, more than 200 exoplanets have been identified orbiting red giant stars \citep{2023Chen, 2025MNRAS.544.1186B, 2026enap....1..520G}, with masses commonly ranging from 0.03 to 10 Jupiter masses (M$_\mathrm{J}$)\footnote{M$_\mathrm{J}$ = 1.89 $\times 10^{27}$ kg $\approx$ $9.54 \times 10^{-4}$ $\mathrm{M}_{\odot}$.} and orbital distances between 0.057 and 3~AU. These findings 
are likely influenced by detection biases inherent to the most effective observational techniques for evolved stars \citep[e.g.,][]{Sato2008, 2015harakawa}.

The evolution of the host star plays a fundamental role in shaping the orbital architecture, survival, and physical properties of its planetary system \citep[e.g.,][]{Villaver2009, Kunitomo2011}. Understanding planetary evolution around evolved stars therefore requires a detailed assessment of the processes that accompany stellar evolution. Numerous studies have investigated the dynamical interactions governing the orbital evolution of planets around solar-type stars \citep[e.g.,][]{Debes2002, MV2012, 2013V}. Models that incorporate tidal effects across different evolutionary stages show that tides can significantly modify planetary orbits and, in some cases, lead to engulfment of close-in companions {\citep[e.g.,][]{Kunitomo2011, MV2012, villaver2014, Madappatt, Ronco}}.

During the red giant branch (RGB) and thermally pulsing asymptotic giant branch (TPAGB) phases, Solar-like stars lose a substantial fraction of their mass through stellar winds \citep[e.g.,][]{Reimers1975, vaswood}. Observations of circumstellar envelopes around evolved stars reveal complex, often non-spherical morphologies attributed to the presence of orbiting companions \citep[e.g.,][]{Decin2020}. This raises the possibility that planets may accrete material from the winds of their host stars \citep{SPMAD, 2017detweiler}, particularly during episodes of intense mass loss such as the RGB and TPAGB phases.

Despite its relevance, the mass accretion history of planets orbiting evolved solar-type stars remains poorly characterized. The first attempt to quantify this was made by \citet{1998Icar..134..303D} who analytically estimated the mass intercepted by planets during the RGB phase of the Sun. This approach to estimate accretion rates was expanded and applied to other planetary remnants and binary companions, including white dwarfs (WD) stars \citep{2006ApJ...652..636D, 2008AJ....135.1785J, 2018MNRAS.473.2871V, 2024MNRAS.527.1014L}. After \citet{1998Icar..134..303D}, a few subsequent studies addressing the accretion by planets orbiting evolving stars have generally relied on the classical Bondi–Hoyle–Lyttleton (BHL) wind accretion model \citep{1939BH, Bondi1944} to estimate mass transfer rates. \citet{Villaver2009} and \citet{SPMAD} examined Jupiter-mass planets under different stellar-wind conditions and found that the total mass accreted is negligible and has little impact on orbital dynamics. However, \citet{SPMAD} noted that accretion luminosity could become significant during phases of high stellar mass-loss rates. Similarly, \citet{2017detweiler} estimated accretion rates using an isothermal Parker wind model but did not provide quantitative predictions.

Several distinct mechanisms can drive mass accretion onto the planet, depending on the system geometry and the host star’s evolutionary stage. When the Roche radius exceeds the condensation radius of the stellar wind, $R_\mathrm{roche} > R_\mathrm{cond}$ \citep{Eggleton1983,2007Hofner} accretion proceeds through gravitational capture of the out-flowing material. This condition is particularly relevant for RGB and AGB stars, whose winds become dust-driven beyond the condensation radius and reach terminal velocity; at this point, the gas is no longer gravitationally confined and cannot be redirected toward the inner Lagrangian point.
If this condition is not satisfied, the wind can instead be focused through the first Lagrangian point, producing Wind Roche-Lobe Overflow (WRLOF; \citealt{2012Mohamed}), which allows more efficient mass transfer. When the stellar radius approaches or exceeds the Roche lobe, $R_\mathrm{star} \gtrsim R_\mathrm{roche}$, Roche-Lobe Overflow (RLOF) occurs and the stellar envelope flows directly through the Lagrangian point. If the planet is engulfed, the system might enter a common-envelope (CE) phase {\citep{Iben1993,Ivanova2013, Nordhaus2013, 2021MNRAS.502L.110C, 2021MNRAS.501..676L, 2021ApJ...915L..34M, 2023ApJ...954..176Y, 2023ApJ...950..128O, 2025PASA...42...27C, 2026ApJ...998..118Y}}, during which the planet experiences drag and may accrete material while spiraling inward.

In this study, we focus on stellar-wind accretion, a process extensively studied in binary systems using the standard BHL accretion model. However, a straightforward application of this prescription yields inaccurate efficiencies when the wind velocity $v_\mathrm{w}$ is comparable to or smaller than the orbital velocity $v_\mathrm{o}$ of the accretor \citep{Boffin2015, Hansen2016}. To overcome this limitation, we employ the modified BHL implementation recently developed by \citet{TejedaToala2025}, which provides a consistent description of wind accretion for objects in circular orbits under $v_\mathrm{w}\lesssim v_\mathrm{o}$ conditions.

The main objective of this study is to quantify the mass accreted by planets via stellar wind during the RGB and TPAGB phases of a 2~$\mathrm{M_\odot}$ host star and to evaluate its impact on planetary dynamical evolution. Additionally, we aim to test the applicability of the modified BHL model in the planetary regime, extending its previous application beyond symbiotic binary systems \citep{Vasquez2024,Maldonado2025,Maldonado2025b}.

This paper is organized as follows. Section \ref{sec2} describes our methodology, which combines the modified accretion formalism, the 2~M$_\odot$ stellar evolution model computed with MESA, and simulations performed with the N-body package REBOUND. Section \ref{sec3} presents our results, Section \ref{sec4} discusses their implications and, Section \ref{sec5} summarizes our main conclusions. 

\section{METHODOLOGY} \label{sec2}

We performed N-body simulations of single-planet systems orbiting an evolving Solar-like star using the REBOUND code \citep{Reinliu2012}, employing the high-accuracy, adaptive IAS15 integrator \citep{Reinspiegel2015}. We incorporate a detailed stellar evolution model at each time-step by interpolating the stellar parameters. 

\begin{figure}[t]
\centering
\includegraphics[width=1.0\linewidth]{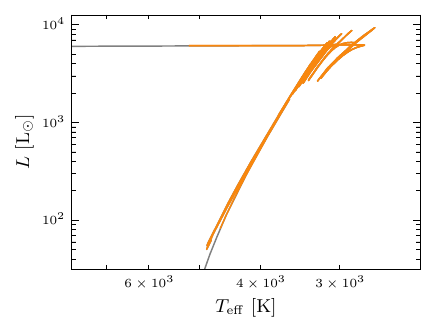}
\captionsetup{justification=justified} 
\caption{Evolutionary track of the 2 M$_\odot$ stellar model used in our simulations. The portion of the segment of the track corresponding to the RGB and TPAGB phases, which are the stages considered in all simulations, is highlighted.}
\label{Fig:1}
\end{figure}

\begin{figure*}[t]
\centering
\includegraphics[width=0.9\linewidth]{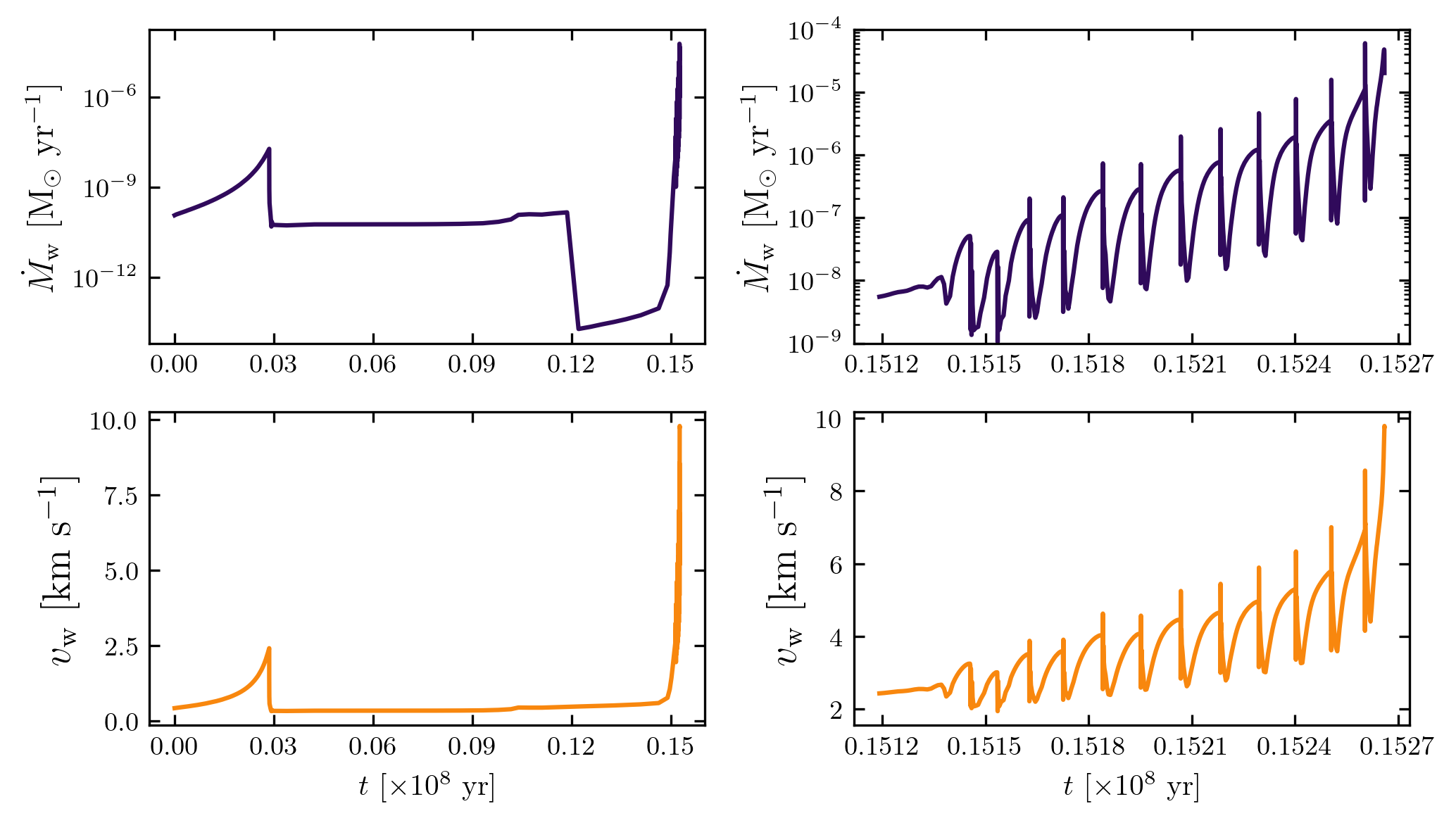}
\captionsetup{justification=justified} 
\caption{Evolution of the stellar mass-loss rate ($\dot{M}_\mathrm{w}$) and wind velocity ($v_\mathrm{w}$) obtained from our 2 M$_\odot$ stellar model. The left panels show the complete evolution of $\dot{M}_\mathrm{w}$ and $v_\mathrm{w}$ through the RGB to the TPAGB phases, while the right panels focus specifically on the TPAGB evolutionary phase.}
\label{Fig:2}
\end{figure*}

The stellar evolution model used in this study was computed with MESA \citep{Paxton2011,2013Paxton,Paxton2015,Paxton2018,Jermyn2023}, using version r15140 \citep{Paxton2019} for a non-rotating star with an initial mass of 2 M$_\odot$. The mass-loss rate  ($\dot{M}_\mathrm{w}$) was prescribed following the Reimers \citep{Reimers1975} and Bl\"{o}cker \citep{Blocker1995} formalisms, adopting efficiency parameters of $\eta_\mathrm{R}=0.5$ during the RGB phase and $\eta_\mathrm{B}=0.1$ during the AGB phase. 
The stellar evolution track is depicted in Fig.~\ref{Fig:1}, where the RGB and AGB phases of the donor star are highlighted. The evolution of the mass-loss rate throughout these phases, as obtained from the MESA calculations, is presented in the top panels of Fig.~\ref{Fig:2}.

Mass accretion by the orbiting planets is included through a wind accretion model. We adopted the model recently developed by \citet{TejedaToala2025}, which introduces a geometric correction to the BHL prescription for binary systems. This correction accounts for the relative geometry and velocity regime of the flow, which is particularly relevant for systems where $v_\mathrm{w} \lesssim v_\mathrm{o}$. 
Recalling that the wind accretion efficiency is defined as the ratio between the mass accretion rate onto the planet and the mass-loss rate from the donor star ($\eta = \dot{M}_\mathrm{acc}/\dot{M}_\mathrm{w}$), it can be expressed for circular orbits as
\begin{equation}
    \eta = \left( \frac{q}{1 + w^2} \right)^2,
\end{equation}
where $q$ and $w$ are the mass and wind dimensionless parameters defined as
\begin{equation}
    q = \frac{m_2}{m_1 +  m_2},
\end{equation}
\begin{equation}
    w = \frac{v_\mathrm{w}}{v_\mathrm{o}}.
\end{equation}
Here, $m_1$ represents the mass of the donor star, $m_2$ is the accretor (the planet) and the orbital velocity, $v_\mathrm{o}$, is defined as
\begin{equation}
   v_\mathrm{o} = \sqrt{\frac{G(m_1 + m_2)}{a}}.
\end{equation}
The wind velocity of the donor star $v_\mathrm{w}$ is computed following the empirical relation derived by \citet{Verbena2011}
\begin{equation}
    v_\mathrm{w} = 0.05 \left( \frac{L_1}{\mathrm{L}_\odot} \cdot \frac{\mathrm{M}_\odot}{m_1} \right)^{0.57} \text{ km s}^{-1},
\end{equation}
where $L_1$ is the luminosity of the donor host star. The resultant evolution of $v_\mathrm{w}$ with time during the RGB and TPAGB phases is illustrated in the bottom panels of Fig.~\ref{Fig:2}.

Using this recently developed accretion model, together with the REBOUND package and MESA,  we performed 18 simulations of planetary systems, each assuming circular orbits. Initial planetary masses were set to $m_\mathrm{2i} = 0.5$, 1, 2.5, 5, 10, and 13~$\mathrm{M}_\mathrm{J}$, with the upper mass limit corresponding to the typical transition between massive planets and brown dwarfs \citep{lecavelier2002}. 

Three initial orbital separations were considered: $a_\mathrm{i} = 5$, 10, and 20 AU. The minimum distance of 5 AU was chosen to ensure that tidal interactions with a $2~\mathrm{M}_\odot$ host star remain negligible \citep[e.g.][]{MV2012,villaver2014} and to avoid entering the WRLOF regime. Also, in all simulations, the condition $R_\mathrm{Roche} > R_\mathrm{cond}$ was satisfied, where $R_\mathrm{Roche} /a = 0.49 \delta^{2/3} / (0.6  \delta^{2/3} + \ln(1 +  \delta^{1/3}))$ with $\delta = m_1 / m_2 $ \citep{1998ApJ...499..853E} and $R_\mathrm{cond} \sim 3 R_\mathrm{1}$ \citep{2007Hofner}, confirming that mass transfer occurred exclusively through stellar wind accretion. As pointed in Section \ref{intro}, accretion can also happen through RLOF or CE evolution, however including these accretion mechanisms should require a full hydrodynamical treatment \citep[e.g.][]{Sawada1992, 2012Passy, 2016Staff}, consequently, those regimes are not considered in the present work.

\section{RESULTS} 
\label{sec3}
\label{sec:results}

\begin{figure*}[t]
\centering
\includegraphics[width=\linewidth]{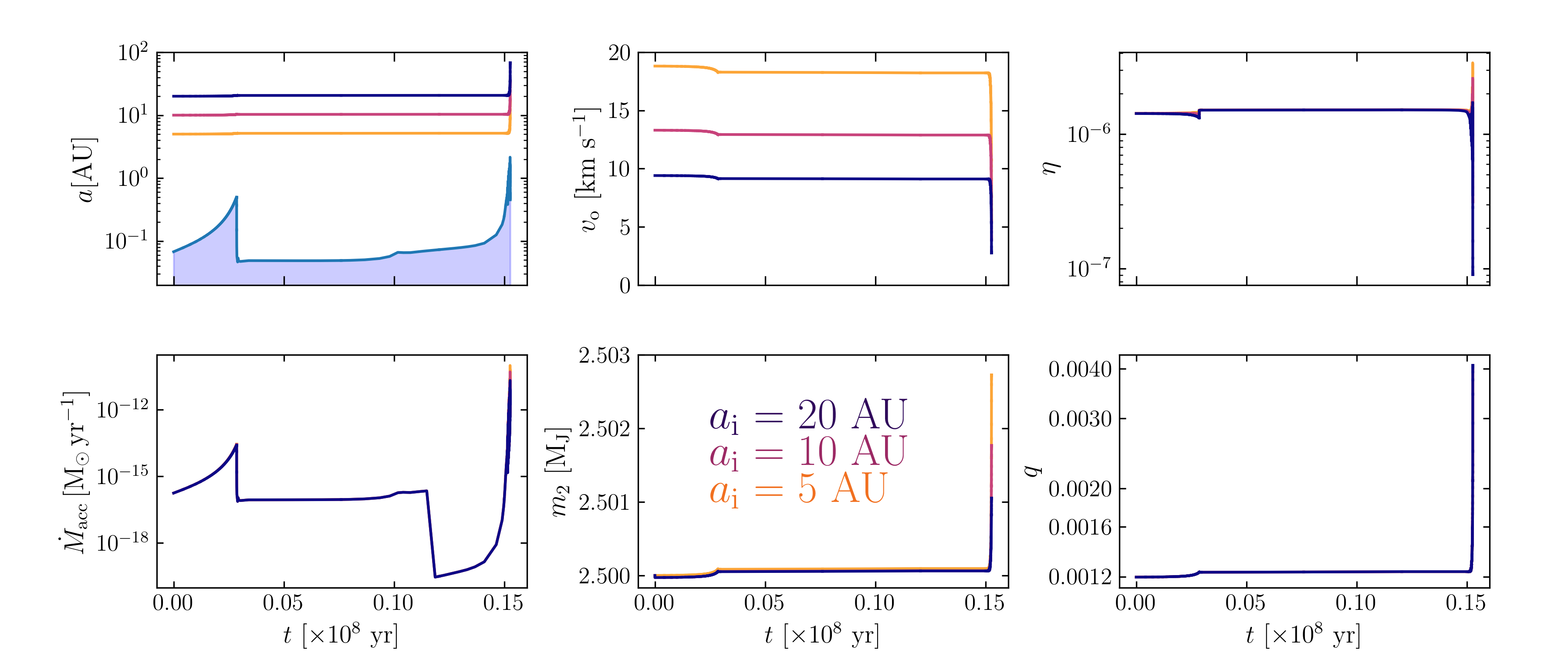}
\captionsetup{justification=justified} 
\caption{Evolution from the RGB phase to the TPAGB phase of all planetary systems with a 2.5 $\mathrm{M}_\mathrm{J}$ planet. The panels show the evolution of the semi-major axis $a$ (top left), orbital velocity $v_o$ (top middle), mass accretion efficiency $\eta$ (top right), mass accretion rate $\dot{{M}}_{\text{acc}}$ (bottom left), planetary mass ${m}_{2}$ (bottom middle), and mass ratio $q$ (bottom right). The solid line over the shaded area indicates the evolution of the stellar radius.}
\label{Fig:3}
\end{figure*}

\begin{figure*}[t]
\centering
\includegraphics[width=\linewidth]{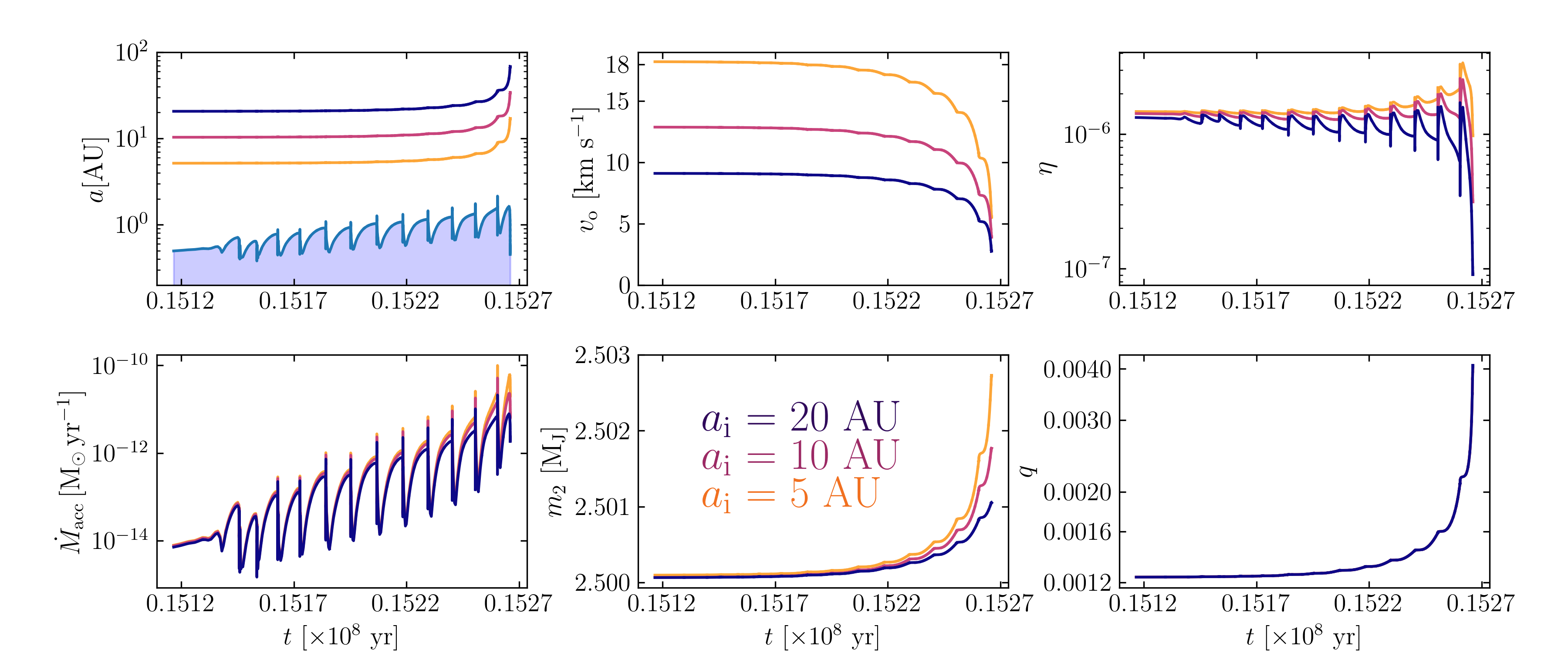}
\caption{Same as Fig.~\ref{Fig:3}, but showing the parameter evolution exclusively during the TPAGB phase.}
\label{Fig:4}
\end{figure*}

\begin{figure*}[t]
\centering
\includegraphics[width=\linewidth]{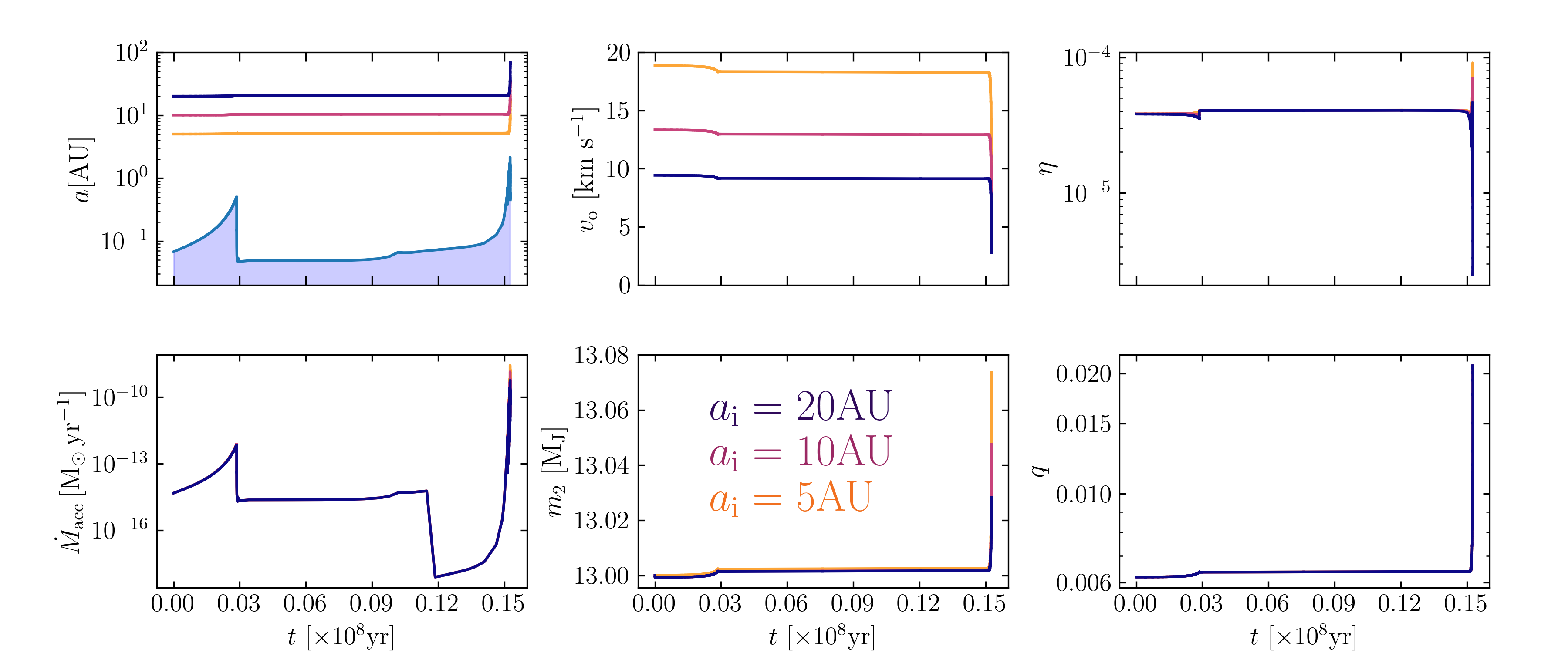}
\caption{ The same panels as Fig. \ref{Fig:3} but showing the evolution of the whole integration time of a 13 $\mathrm{M}_\mathrm{J}$ planet. }
\label{Fig:5}
\end{figure*}

\begin{figure*}[t]
\centering
\includegraphics[width=\linewidth]{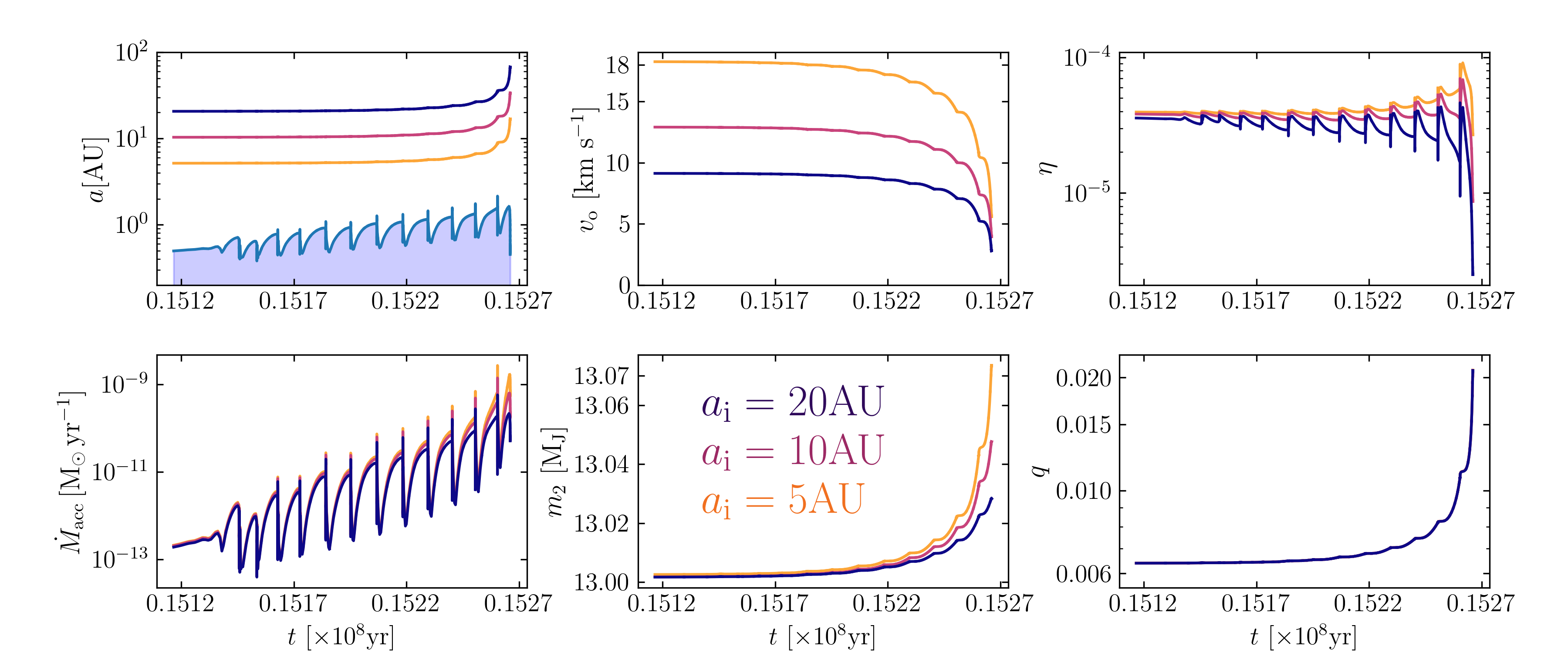}
\caption{Same as Fig.~\ref{Fig:5}, but showing the parameter evolution exclusively during the TPAGB phase.}
\label{Fig:6}
\end{figure*}

Table \ref{tab:1} presents the initial semi-major axes, the initial masses of each simulated planet, their accreted mass, and the percentage of the planet’s initial mass that was accreted by it. Data in this table shows that the highest mass accretion occurs in planetary systems with the smallest initial orbital separation, $a_\mathrm{i} = 5$ AU, and that the total accreted mass increases with the planet's initial mass. Consequently, the planetary system with the closest and most massive planet ($a_\mathrm{i} = 5$ AU and $m_{\mathrm{2i}} = 13$ $\mathrm{M}_\mathrm{J}$) exhibits the greatest mass accretion, approximately $\sim$0.56$\%$ of the planet's initial mass, which is in agreement with the estimation from \citet{1998Icar..134..303D}, that a planet may accrete less than one percent of its mass during the post main sequence evolution of its host star. 

\begin{table}
    \centering
    \begin{tabular}{c|c|c|c}
        \toprule
        $a_\mathrm{i}$ (AU) & $m_\mathrm{2i}$ ($\mathrm{M}_\mathrm{J}$) & $m_\mathrm{acc}$ ($\mathrm{M}_\mathrm{J}$) & $\Delta m_2$ (\%) \\
        \midrule
        \multirow{6}{*}{5}   & 0.5 & 0.0001  &  0.02 \\
                            & 1  & 0.00043  & 0.04 \\
                            & 2.5 & 0.0027  & 0.1 \\
                             & 5  & 0.01090  & 0.21 \\
                            & 10 & 0.0435 & 0.43 \\
                             & 13 & 0.0734 & 0.56 \\

        \midrule
        \multirow{6}{*}{10}  & 0.5 & 0.00007 & 0.01 \\
                            & 1  & 0.00028  & 0.02 \\
                             & 2.5 & 0.00176 & 0.07 \\
                             & 5  & 0.00706  & 0.14 \\
                             & 10 &  0.02819 & 0.28 \\
                             & 13 & 0.04760 &  0.36 \\
        \midrule
        \multirow{6}{*}{20}  & 0.5 &   0.00004 &  $8 \times 10^{-3}$ \\
                             & 1  & 0.00016  & 0.01 \\
                             & 2.5 & 0.00105 & 0.04 \\
                             & 5  & 0.00419  & 0.08 \\
                             & 10 & 0.01677 &  0.16 \\
                             & 13 & 0.02831 &  0.21 \\
        \bottomrule
    \end{tabular}
    \captionsetup{justification=justified} 
    \caption{Initial parameters of the planets in the simulations, including the initial semi-major axis ($a_\mathrm{i}$) and initial planetary mass ($m_\mathrm{2i}$), as well as the mass accreted by each planet ($m_\mathrm{acc}$) and the percentage of the planet’s initial mass that was accreted by it, that is, $\Delta m_2 = (m_\mathrm{acc}/m_\mathrm{2i})~\times$ 100.}
    \label{tab:1}
\end{table}

Figures \ref{Fig:3} and \ref{Fig:5} illustrate the time evolution of orbital and accretion-related parameters from the RGB to the TPAGB phase for planetary systems hosting 2.5 and 13 $\mathrm{M}_{\mathrm{J}}$ planets, respectively, with initial semi-major axes of 5, 10, and 20 AU. The parameters shown in both figures include the semi-major axis ($a$), orbital velocity ($v_\mathrm{o}$), mass accretion efficiency ($\eta$), mass accretion rate ($\dot{M}_\mathrm{acc}$), planetary mass ($m_{2}$), and mass ratio ($q$). In these plots, the shaded region represents the stellar radius, while each curve corresponds to different initial orbital separations, as specified in the legend. Throughout most of the evolution, $a$, $v_\mathrm{o}$, $\eta$, $m_2$, and $q$ remain nearly constant, with noticeable changes occurring only toward the end of the TPAGB phase, where the orbital expansion of the accretor, driven by stellar mass loss, leads to corresponding changes in the other parameters. 

\begin{figure*}[t]
\centering
\includegraphics[width=\linewidth]{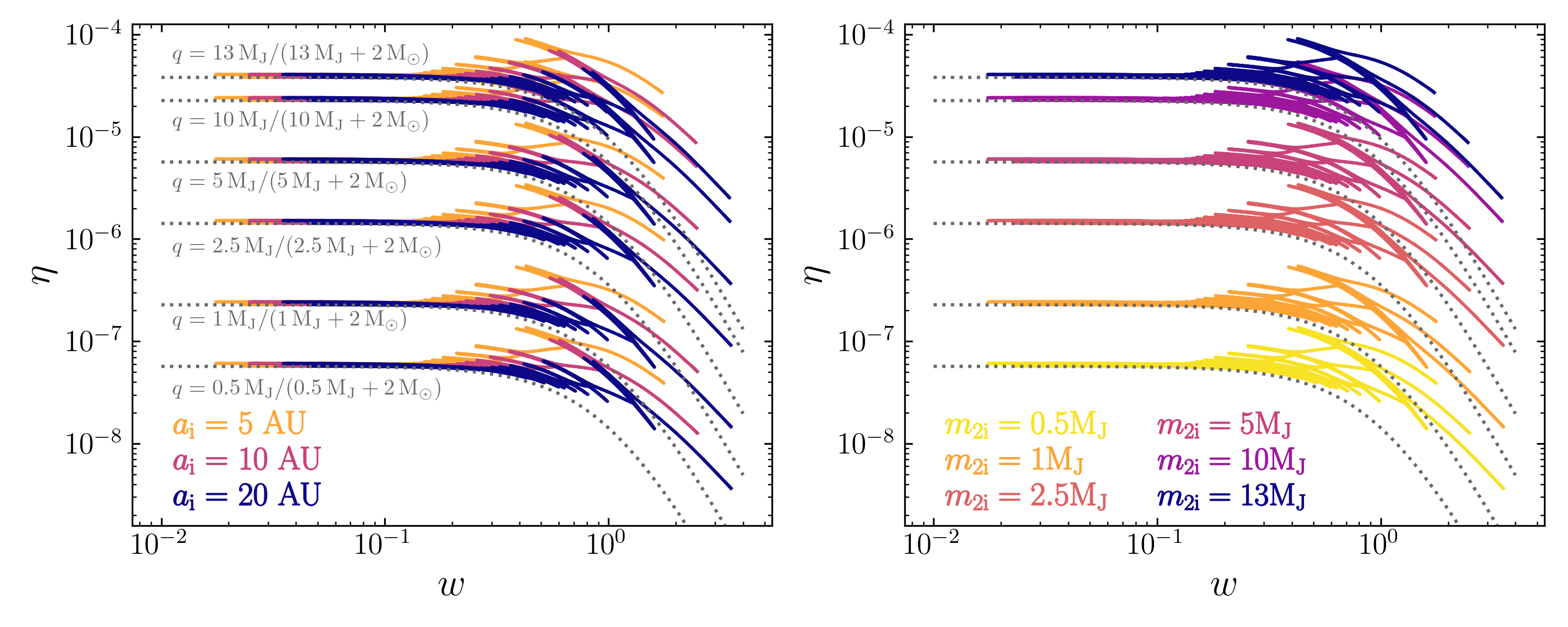}
\captionsetup{justification=justified} 
\caption{Evolution of the accretion efficiency $\eta$ as a function of the velocity ratio for all planetary systems. The left panel shows curves corresponding to the initial semi-major axes $a_\mathrm{i}$, while the right panel shows curves for different initial planetary masses $m_\mathrm{2i}$. Dotted lines correspond to analytical estimates of $\eta$ for constant mass ratio $q$ values.}
\label{Fig:7}
\end{figure*}

\begin{figure*}[ht!]
\centering
\includegraphics[width=\linewidth]{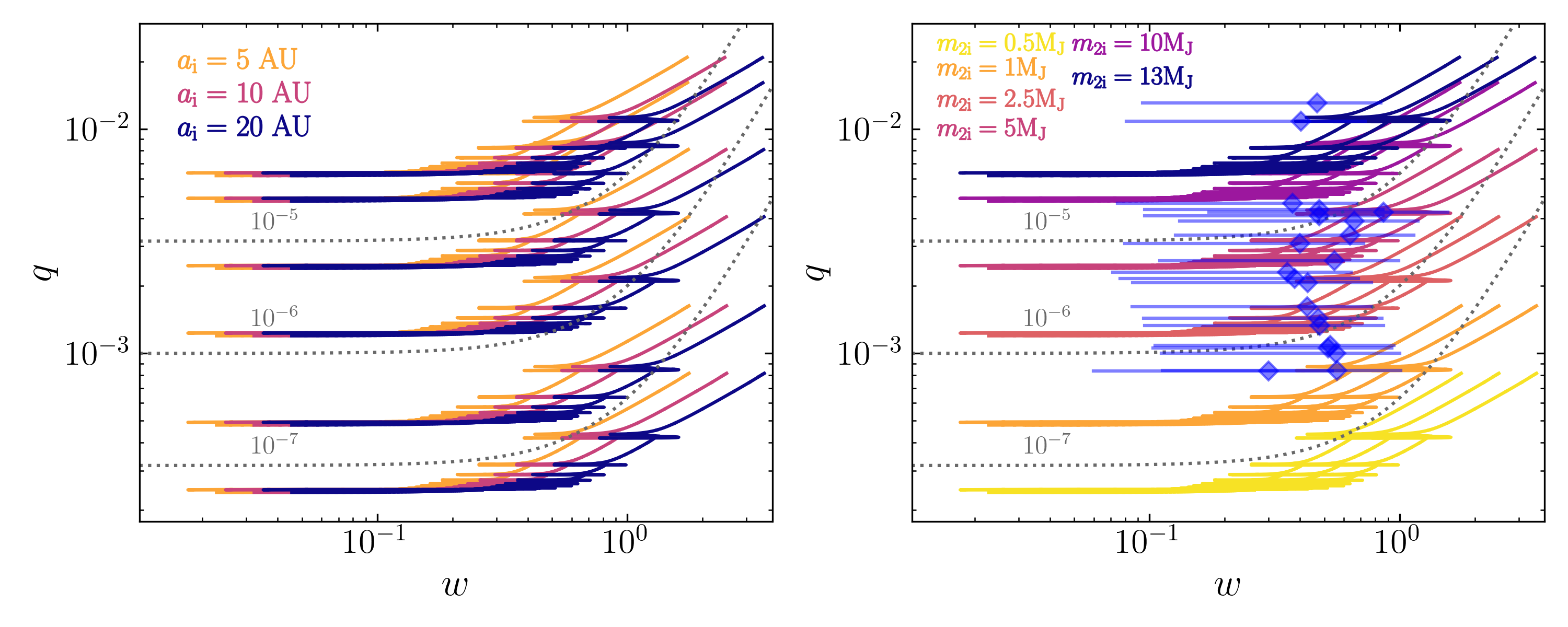}
\captionsetup{justification=justified} 
\caption{Relationship between $w$ and $q$ for all planetary systems. The left panel shows curves corresponding to different initial semi-major axes $a_\mathrm{i}$, while the right panel shows curves for different initial planetary mass $m_\mathrm{2i}$. Dotted lines represent contours of constant values of $\eta$. Diamond symbols mark the observed exoplanet systems listed in Table \ref{tab:2}, for which $v_\mathrm{w}$ was calculated assuming a wind velocity of $v_\mathrm{w}$ = 10 $\pm$ 8 km s$^{-1}$.}
\label{Fig:8}
\end{figure*}

Figures \ref{Fig:4} and \ref{Fig:6} present the same set of parameters for the same planetary systems as Figures \ref{Fig:3} and \ref{Fig:5}, but focusing on the TPAGB phase of the 2 $\mathrm{M}_{\odot}$ star. During this stage, more pronounced variations are observed in $a$, $v_\mathrm{o}$, $\eta$, $m_2$, and $q$ reflecting the stronger dynamical and accretional responses to the enhanced mass loss. In particular, the upper-right and lower-left panels, which display $\eta$ and $\dot{M}_\mathrm{acc}$, respectively, clearly illustrate the pulsating behavior of the 2 $\mathrm{M}_{\odot}$ host star. During the TPAGB phase, the accretion rate ranges from $\sim$$10^{-15}$ to $\sim$$10^{-10}$ $\mathrm{M}_{\odot}$ yr$^{-1}$ for 2.5 $\mathrm{M}_{\mathrm{J}}$ planets and from $\sim$$10^{-14}$ to $\sim$$10^{-9}$ $\mathrm{M}_{\odot}$ yr$^{-1}$ for 13 $\mathrm{M}_{\mathrm{J}}$ planets. The remaining simulations exhibit similar qualitative behavior to those shown in Figures \ref{Fig:3}$-$\ref{Fig:6}; therefore, we present only the results of the simulations for 2.5 and 13 $\mathrm{M}_{\mathrm{J}}$ planets. Note, that the dimensionless mass parameter $q$, is nearly identical in Figures \ref{Fig:3}$-$\ref{Fig:6}, as the total mass accreted is negligible compared to the planet’s initial mass. 

The evolution of the accretion efficiency $\eta$ as a function of the velocity ratio $w$ for all simulations is shown in Figure \ref{Fig:7}. The left panel differentiates planets according to their initial semi-major axes, while the right panel distinguishes them by their initial planetary masses. In each panel, the solid lines represent the computed accretion efficiencies from the simulations, while the dotted lines correspond to analytical contours of $\eta$ computed for constant mass-ratio values $q$. At the beginning of the simulations, the systems follow these analytical contours closely,  reflecting the consistency between the numerical results and the analytical predictions under relatively stable orbital and wind conditions. Throughout most of the evolution, $w < 1$, as the orbital velocity remains higher than the wind velocity. As the systems approach the TPAGB phase, the combined effects of thermal pulses, the progressive increase in wind velocity and the orbital expansion of the planet lead to deviations from the constant-$q$ contours. During this stage, the stellar mass loss causes $m_1$ to decrease significantly, while the accretor's mass grows slightly with each thermal pulse, resulting in a gradual increase in $q$. This raise in $q$ partially offsets the decline in $\eta$ caused by the increasing $w$, producing a mild overall increase in the accretion efficiency.  This behavior illustrates how the joint evolution of the mass ratio and wind-orbital velocity balance governs the accretion efficiency during the late stages of stellar evolution. 

Figure \ref{Fig:8} presents the evolution of the 18 simulated planetary systems in the $q-w$ parameter space. The dotted lines represent contours of constant accretion efficiency $\eta$. In the left panel, each curve corresponds to a different initial semi-major axes, while in the right panel the curves distinguish systems with different initial planetary masses. In general, both $q$ and $w$ increase over time as a consequence of the stellar mass loss and the gradual accretion of material by the planetary companion.  
The right panel also includes the observed exoplanet systems listed in Table \ref{tab:2}, shown as diamond symbols. For these systems, $w$ was calculated assuming a wind velocity of $v_\mathrm{w}$ = 10 $\pm$ 8 km s$^{-1}$, enabling a direct comparison with the simulations. These observed systems occupy the same region of the $w -q$ space as the high-mass simulated planets, indicating good agreement between the models and the observations. The selection criteria and physical properties of these observed systems are described in detail in Section~\ref{sec4}, where we compare them more extensively with the simulation results.

\begin{table*}
\centering
\setlength{\tabcolsep}{0.75\tabcolsep}  
\renewcommand{\arraystretch}{1.0}
\captionsetup{justification=justified} 
\caption{Properties of the sample of 21 exoplanets orbiting red giant stars from \citet{2023Chen}, with semi-major axes $a\geq2.5$~AU and planetary masses $M_{\mathrm{pl}} \geq 1~\mathrm{M}_\mathrm{J}$, together with the properties of their host stars. Exoplanets marked with an asterisk belong to two-planet systems.}
\begin{tabular}{lcccccccccccc}
    \toprule
    Planet name & $m_{\mathrm{2}}$  & $a$  & $e$ & $R_{\mathrm{2}}$  & $T_{\mathrm{eq}}$ & $L_{\mathrm{eq}}$  & $m_{\text{1}}$  & $T_{\text{1}}$ & $R_{\text{1}}$ & $w$ & $q$ & $L_{\mathrm{acc}}$/$L_{\mathrm{eq}}$ \\
      & ($\mathrm{M}_\mathrm{J}$) & (AU) &   & ($\mathrm{R}_\mathrm{J}$) &  (K) & (erg s$^{-1}$) & ($\mathrm{M}_{\odot}$) & (K) &  ($\mathrm{R}_\odot$) &  &  \\
    \midrule
    18 Del b & 10.3 & 2.6 & 0.024 & 1.020 & 15.916 & $2.42 \times 10^{21}$ & 2.10 & 4980 & 8.5 & 0.3727 & 0.0047 & $2.35 \times 10^{7}$ \\
    81 Cet b & 5.3 & 2.5 & 0.037 & 1.061 & 20.752 & $7.58 \times 10^{21}$ & 1.23 & 4734 & 11.21 & 0.4777 & 0.0041 & $2.75\times 10^{6}$ \\
    BD+49 828 b & 1.6 & 4.2 & 0.35  & 1.140 & 10.589 & $5.93\times 10^{20}$ & 1.52 & 4943 & 9.20 & 0.5578 & 0.0010 & $5.71 \times 10^{}$ \\
    HD 219415 b & 1.0 & 3.2 & 0.4 & 1.173 & 6.286 & $ 7.79\times 10^{19}$ & 1.138 & 4815 & 4.27 & 0.5628 & 0.0008 & $1.83 \times 10^{6}$ \\
    HD 238914 b & 6.0 & 5.7 & 0.56 & 1.053 & 12.009 & $8.37\times 10^{20}$ & 1.47 & 4769 & 14.68 & 0.6598 & 0.0039 & $ 2.32\times 10^{7}$ \\
    HIP 114933 b & 1.94 & 2.84 & 0.21 & 1.127 & 8.749 & $2.70 \times 10^{20}$ & 1.39 & 4823 & 5.27 & 0.4796 & 0.0013 & $ 2.81\times 10^{6}$ \\
    HIP 56640 b & 3.67 & 3.73 & 0.118 & 1.085 & 6.162 & $6.15 \times 10^{19}$ & 1.04 & 4769 & 4.93 & 0.6348 & 0.0034 & $ 1.42\times 10^{8}$ \\
    HIP 65891 b & 6.0 & 2.81 & 0.128 & 1.053 & 15.720 & $2.45 \times 10^{21}$ & 2.490 & 5060 & 8.93 & 0.3563 & 0.0023 & $ 3.19\times 10^{6}$ \\
    HIP 67537 b & 11.09 & 4.91 & 0.586 & 1.015 & 8.250 & $ 1.73 \times 10^{20}$ & 2.41 & 4985 & 8.31 & 0.4782 & 0.0044 & $ 3.01\times 10^{8}$  \\
    HIP 97233 b & 20.0 & 2.55 & 0.63 & 0.980 & 9.4737 & $2.80 \times 10^{20}$ & 1.74 & 5020 & 4.92 & 0.4042 & 0.0109 & $ 2.2\times 10^{9}$ \\
    HD 111591 b & 4.4 & 2.5 & 0.26 & 1.073 & 16.667 & $3.22 \times 10^{21}$ & 1.94 & 4871 & 8.75 & 0.3807 & 0.0022 & $ 1.54\times 10^{6}$  \\
    HD 120084 b & 4.50 & 4.3 & 0.483 & 1.072 & 12.267 & $ 9.44\times 10^{20}$ & 2.661 & 4892 & 11.03 & 0.4265 & 0.0016 & $ 2.94\times 10^{6}$  \\
    HD 125390 b & 22.2 & 3.16 & 0.597 & 0.974 & 7.909 & $1.34 \times 10^{20}$ & 1.60 & 4882 & 5.23 & 0.4687 & 0.0131 & $ 7.24\times 10^{9}$  \\
    HD 14067 b & 7.79 & 3.4 & 0.65 & 1.037 & 14.909 & $1.92 \times 10^{21}$ & 2.4 & 4815 & 10.76 & 0.3990 & 0.0031 & $ 9.58\times 10^{6}$ \\
    HD 142245 b & 3.069 & 2.78 & 0.0 & 1.096 & 7.725 & $1.55 \times 10^{20}$ & 3.50 & 4922 & 4.46 & 0.2991 & 0.0008 & $ 3.36\times 10^{6}$ \\
    HD 145934 b & 2.28 & 4.6 & 0.078 & 1.116 & 7.105 & $1.15 \times 10^{20}$ & 2.058 & 4823 & 6.93 & 0.5017 & 0.0011 & $ 6.18\times 10^{6}$ \\
    HD 1605 c$\ast$ & 3.62 & 3.58 & 0.099 & 1.086 & 5.292 & $3.35\times 10^{19}$ & 1.33 & 4933 & 3.92 & 0.5501 & 0.0026 & $1.6\times 10^{8}$  \\
    HD 75784 c$\ast$  & 5.64 & 8.4 & 0.489 & 1.057 & 1.928 & $5.60\times 10^{17}$  & 1.26 & 4867 & 3.40 & 0.8650 & 0.0043 & $3.4\times 10^{10}$  \\
    HD 86950 b & 3.6 & 2.72 & 0.17 & 1.086 & 17.91 & $4.40 \times 10^{21}$ & 1.66 & 4838 & 10.29 & 0.4293 & 0.0021 & $ 8.18\times 10^{5}$  \\
    HD 94834 b & 1.26 & 2.74 & 0.14 & 1.157 & 7.832 & $1.82 \times 10^{20}$ & 1.11 & 4798 & 4.57 & 0.5272 & 0.0011 & $ 1.70\times 10^{6}$  \\
    kap CrB b & 2.0 & 2.65 & 0.04 & 1.125 & 8.356 & $2.24 \times 10^{20}$ & 1.33 & 4840 & 4.68 & 0.4736 & 0.0014 & $ 4.07\times 10^{6}$\\
    \bottomrule
\end{tabular}
\label{tab:2}
\end{table*}

\section{DISCUSSION} \label{sec4}
\label{sec:discussion}

\begin{figure*}[ht]
\centering
\includegraphics[width=\linewidth]{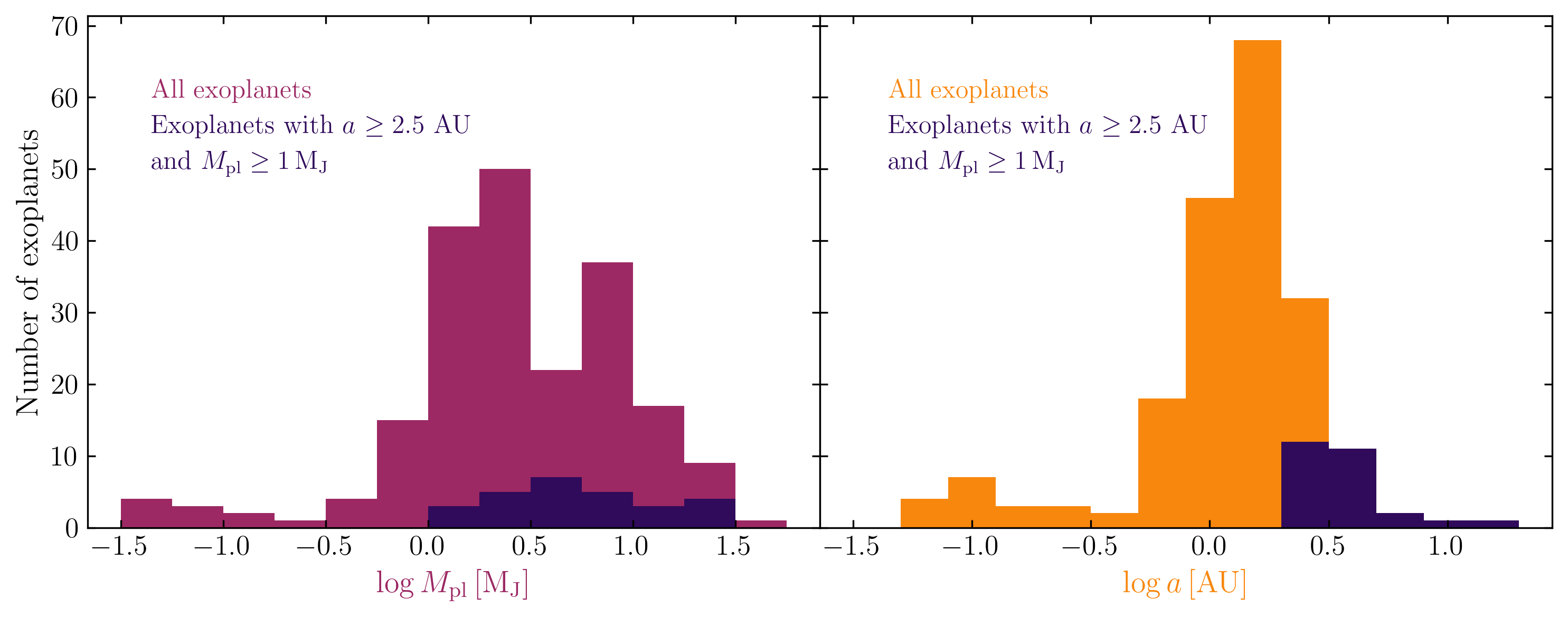} 
\captionsetup{justification=justified} 
\caption{Mass and semi-major axis distributions of the 210 exoplanets orbiting RGS from the sample of \citet{2023Chen}. The exoplanets listed in Table \ref{tab:2} are highlighted as darker shaded bars in both panels.}
\label{Fig:9}
\end{figure*}

To compare our simulation results with observations, we conducted a literature search for exoplanets orbiting evolved stars on the RGB and AGB phases. Although no confirmed AGB host stars were identified, we adopted the recent catalogue by \citet{2023Chen}, which compiles an extensive and homogeneous dataset of 210 exoplanets orbiting RGB stars, representing one of the most comprehensive samples currently available. The distributions of their planetary masses and semi-major axes are shown in Figure \ref{Fig:9}.

From this catalog, we selected a subset of 21 exoplanets with semi-major axes $a \geq 2.5$~AU, since planets at smaller orbital distances are likely to be engulfed by their host star during the RGB phase of its host star \citep[e.g.,][]{villaver2014}. We further restricted the sample to systems where the estimated Roche radius satisfies $R_\mathrm{roche} >$ $R_\mathrm{cond}$ $\sim$ $3\,R_{\mathrm{star}}$, consistent with the typical dust condensation radius for giant stars \citep{2007Hofner}. At this distance, the stellar wind becomes dust-driven and reaches terminal velocity. This ensures that mass accretion onto the planets occurs purely through the capture of a freely expanding stellar wind, rather than through WRLOF \citep{2012Mohamed} or other enhanced accretion mechanisms. The parameters of these 21 exoplanets and their host stars are listed in Table \ref{tab:2}.

Of the 21 selected exoplanets, 19 belong to single-planet systems, while the remaining two are in two-planet systems. In both of these multi-planet systems, the outer and inner planets satisfy the condition $R_\mathrm{roche}>R_\mathrm{cond}\sim3\,R_{\mathrm{star}}$. However, the two inner planets do not meet the $a \geq 2.5$~AU criterion and were therefore excluded in Table \ref{tab:2}. These systems are nevertheless included and discussed in Section~\ref{sec4.6}, since the condition $R_\mathrm{roche} > R_\mathrm{cond}$ indicates that wind accretion could be taking place, even though their relatively short orbital separations suggest that planetary engulfment could occur in the near future.

\subsection{Accretion and equilibrium luminosities} \label{sec4.1}  

To characterize the luminosity sources of the simulated planets, we computed both the bolometric luminosity produced by the accretion process ($L_\mathrm{acc}$) and the equilibrium luminosity ($L_\mathrm{eq}$). This comparison allows us to quantify the relative contribution of each component to the total luminosity of the planets. The accretion luminosity  was calculated as \citep[e.g.,][]{Shakura1973, Pringle1981} :

\begin{center} 
\begin{equation} \label{eq6}
    L_{\mathrm{acc}} = \frac{1}{2} \frac{G m_2}{ R_2} \dot{M}_\mathrm{\mathrm{acc}},
\end{equation}
\end{center}
where  $ R_2 $ is the radius of the planet and  $ \dot{M}_\mathrm{acc} $ = $\eta \dot{M}_\mathrm{w}$, is the mass accretion rate obtained from our simulations. The planetary radius was estimated using  empirical mass-radius relationship \citep[see][]{Chen2017, muller2024}
\begin{center}
\begin{equation} \label{eq7}
    R_2 = (18.6 \pm 6.7)  m_2^{-0.06 \pm 0.07}.
\end{equation}
\end{center}

The equilibrium luminosity of each simulated planet was computed as
\begin{equation} \label{eq8}
L_\mathrm{eq} = 4 \pi R_2^2 \sigma T_\mathrm{eq}^4, 
\end{equation}
where the equilibrium temperature is defined as
\begin{center}
\begin{equation} \label{eq9}
    T_\mathrm{eq} = T_\mathrm{1} \sqrt{\frac{R_\mathrm{1}}{2a}} (1-A_\mathrm{B})^{1/4},
\end{equation}
\end{center}
with $R_1$ and $T_1$ denoting the radius and effective temperature of the host star, respectively, and $A_\mathrm{B}$ is the Bond albedo.

For all simulated planets, we adopted $A_\mathrm{B} = 0.503$, corresponding to Jupiter’s value \citep{Li2018}, since the albedo can not be directly constrained from our models. Using the same assumptions, we also estimated $L_\mathrm{eq}$ for the observed exoplanets listed in Table~\ref{tab:2}, applying equations~(\ref{eq8}) and~(\ref{eq9}). Because the albedoes of the observed exoplanets are uncertain, we included an additional uncertainty of $\pm 0.4$ in $A_\mathrm{B}$. 

\begin{figure*}
\centering
\includegraphics[width=\linewidth]{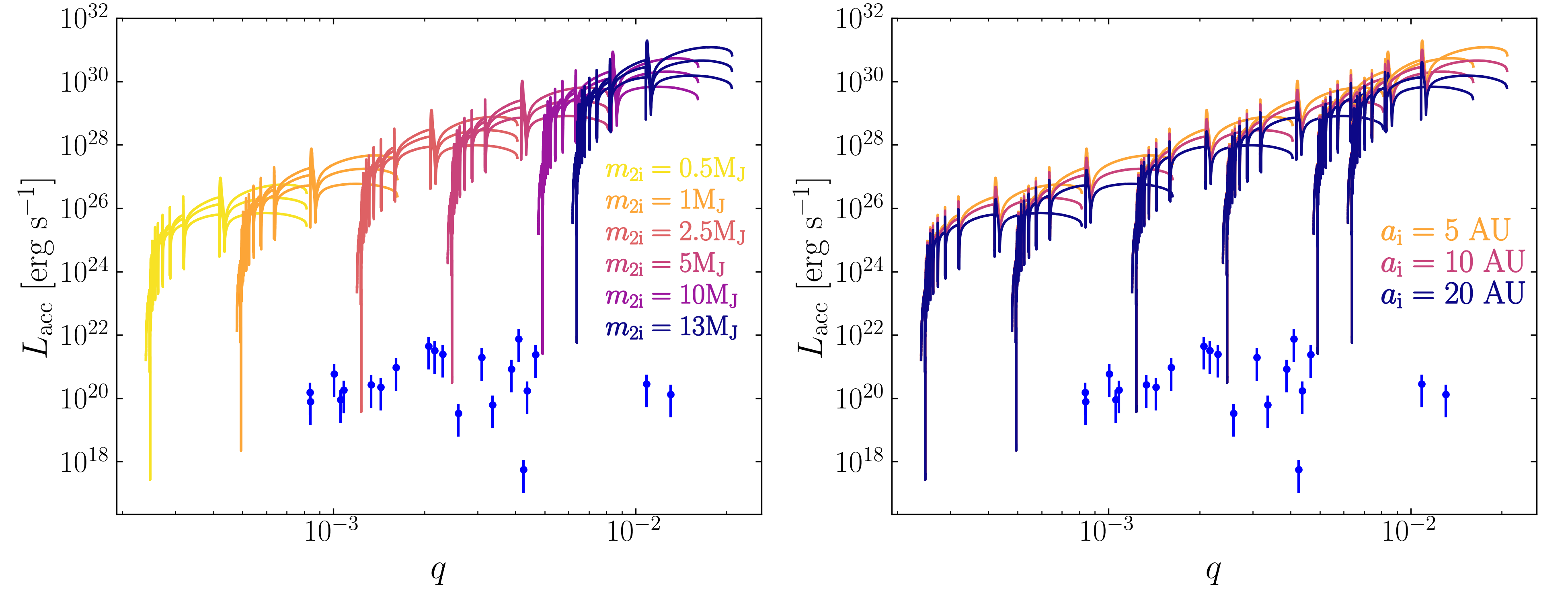}
\captionsetup{justification=justified} 
\caption{Evolution of the accretion luminosity of the simulated planets and the mass ratio $q$. The left panel shows curves grouped by initial planetary mass $m_\mathrm{2i}$, while the right panel shows curves grouped by initial semi-major axis $a_\mathrm{i}$. Dots denote the inferred equilibrium luminosities of the observed planets listed in Table~\ref{tab:2} for $A_\mathrm{B} = 0.503$, and the vertical bars indicate the corresponding luminosity ranges obtained by varying the Bond albedo within $A_\mathrm{B} = 0.503 \pm 0.4$.}
\label{Fig:10}
\end{figure*}

\begin{figure*} 
\centering
\includegraphics[width=\linewidth]{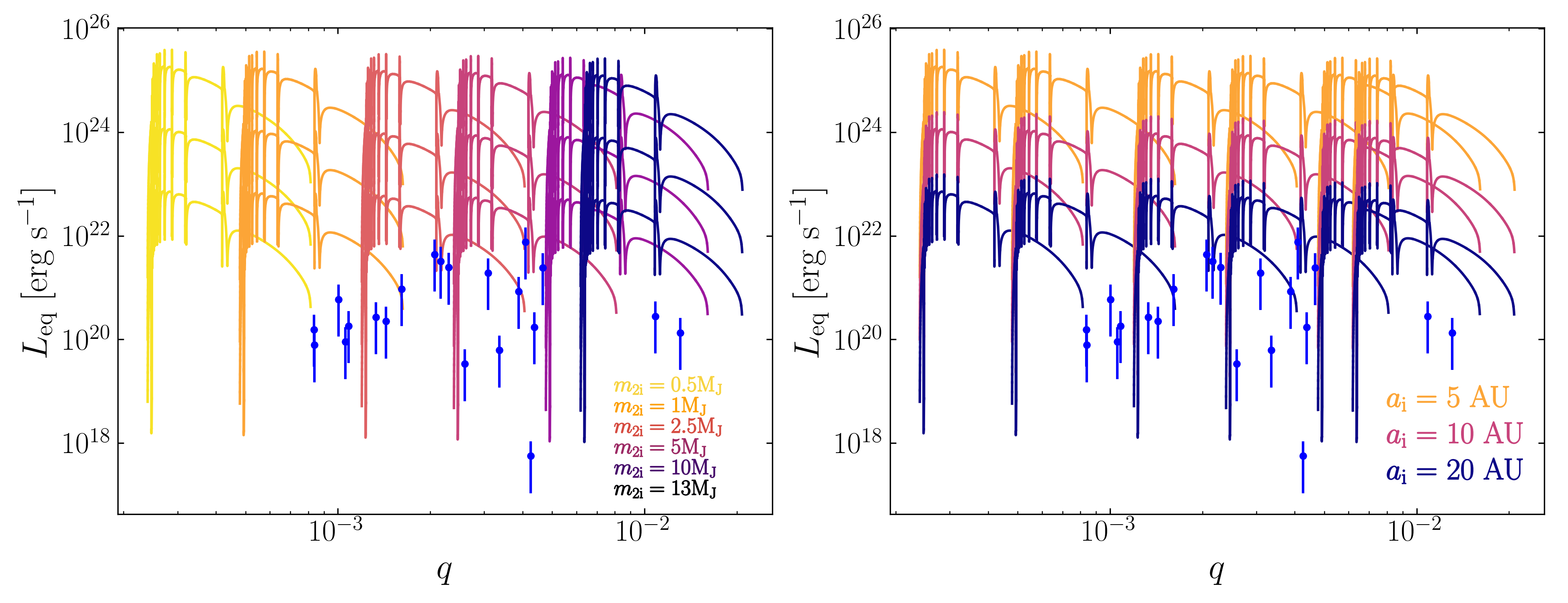}
\caption{Same as Fig.~\ref{Fig:10}, but showing the equilibrium luminosity of the simulated planets.}
\label{Fig:11}
\end{figure*}

The evolution of the accretion luminosity as a function of $q$ is presented in Fig.~\ref{Fig:10}. The left panel distinguishes the simulations by their initial planetary masses $m_\mathrm{2i}$, while the right panel differentiates them by their initial semi-major axes $a_\mathrm{i}$. The dots in both panels represent the equilibrium luminosities estimated for the 21 detected exoplanets of Table~\ref{tab:2}. The simulated planets exhibit accretion luminosities that are generally higher than the expected equilibrium luminosities of the observed exoplanets, which occupy similar ranges of $a_\mathrm{i}$ and $m_{\mathrm{2i}}$. This suggests that, in principle, if the luminosity of an accreting planet is detected, that luminosity would correspond to the accretion luminosity of the planet orbiting an evolved star, assuming that the accretion process and the properties of the stellar wind (mass loss rate and wind velocity) are well understood.

Figure \ref{Fig:11} shows the equilibrium luminosities of the simulated planets as a function of $q$, adopting the same panel organization as Figure~\ref{Fig:10}. The estimated equilibrium luminosities of the detected exoplanets are shown as dots and are in reasonable agreement with our lower predicted values (equilibrium luminosities during the RGB phase). This agreement indicates that the evolution of $L_\mathrm{eq}$ in our simulations is consistent with that expected for known exoplanets orbiting evolved stars.

\begin{figure*}  
\centering
\includegraphics[width=\linewidth]{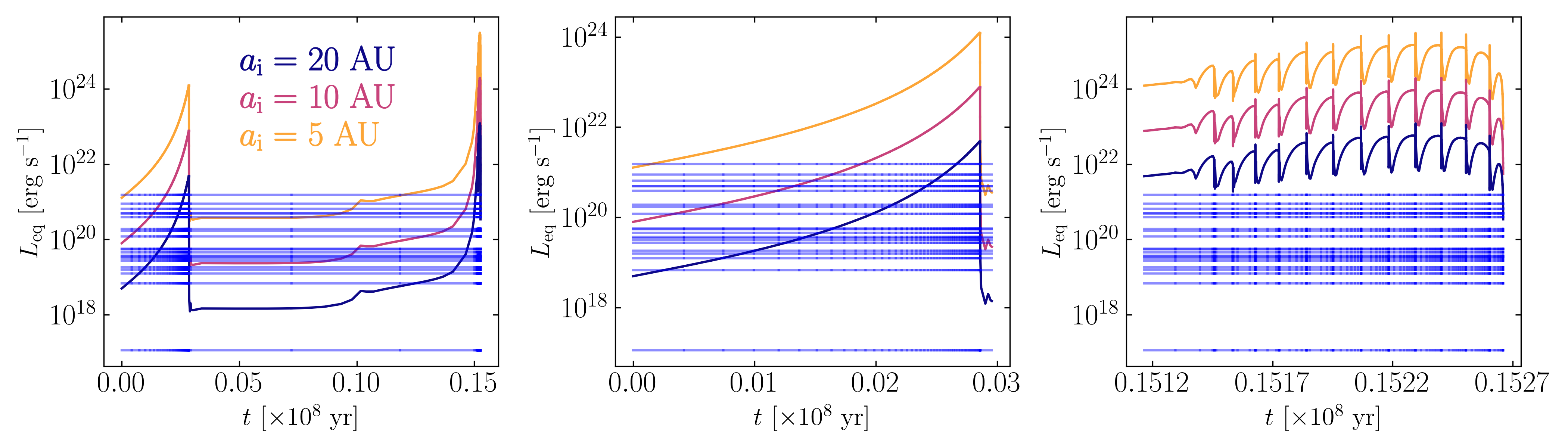}
\captionsetup{justification=justified} 
\caption{Evolution of the median equilibrium luminosity of the simulated planets, binned by semi-major axis. The left panel shows the entire evolution, the central panel corresponds to the RGB phase of the host star, and the right panel to the TPAGB phase. The estimated equilibrium luminosities of the observed planets listed in Table \ref{tab:2} are shown as horizontal solid lines extended across the time axis for reference.}
\label{Fig:12}
\end{figure*}

The median equilibrium luminosity ($<L_\mathrm{eq}>$) of the simulated planets, binned by semi-major axis, is shown in Fig. \ref{Fig:12}. The left panel presents the complete evolution, while the central and right panels isolate the RGB and TPAGB phases, respectively. The equilibrium luminosities estimated from observed exoplanets are indicated by horizontal solid lines extended across the time axis for reference. This figure, confirms that the simulated planets display equilibrium luminosities comparable to those of the observed exoplanets during the RGB and the early TPAGB phase.

\begin{figure*} 
\centering
\includegraphics[width=\linewidth]{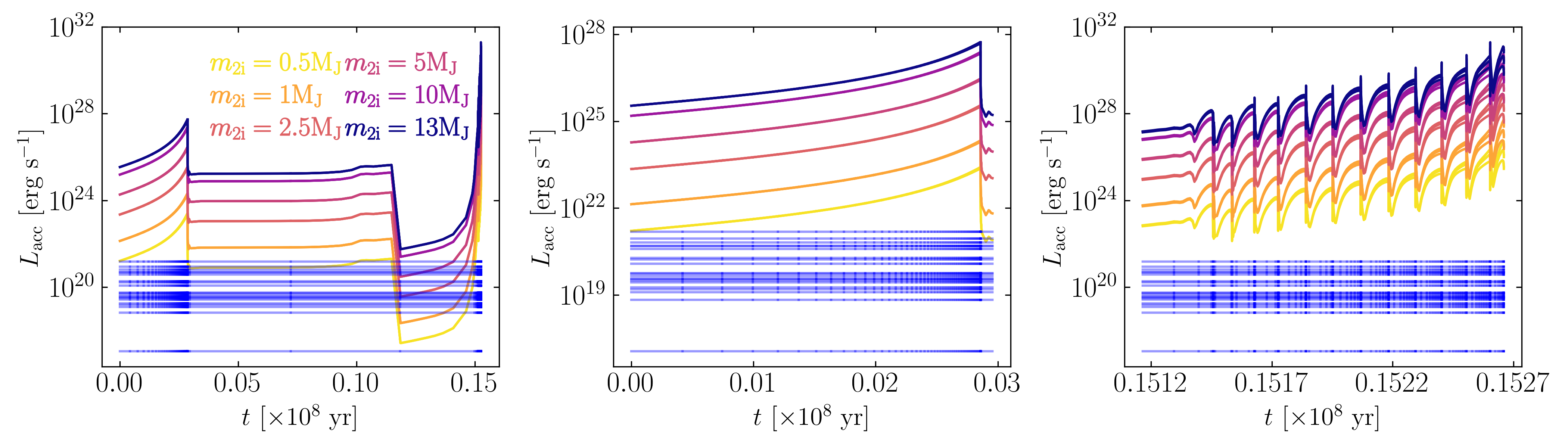}
\captionsetup{justification=justified} 
\caption{Same as Fig.~\ref{Fig:12}, but showing the accretion luminosity of the simulated planets, with different colors indicating the planetary masses.}
\label{Fig:13}
\end{figure*}

Fig.~\ref{Fig:13} shows the accretion luminosity ($L_{\mathrm{acc}}$) of the simulated planets throughout the evolution of the $2\mathrm{M}_\odot$ stellar model, together with the equilibrium luminosities ($L_{\mathrm{eq}}$) of the observed exoplanets. The figure is presented in the same format as Fig.~\ref{Fig:12}, but with different colors indicating the planetary masses. Nearly all $L_\mathrm{eq}$ values of the observed exoplanets lie below the accretion luminosities of the simulations during both the RGB and TPAGB phases (central and right panels).

\begin{figure*}[t] 
\centering
\includegraphics[width=\linewidth]{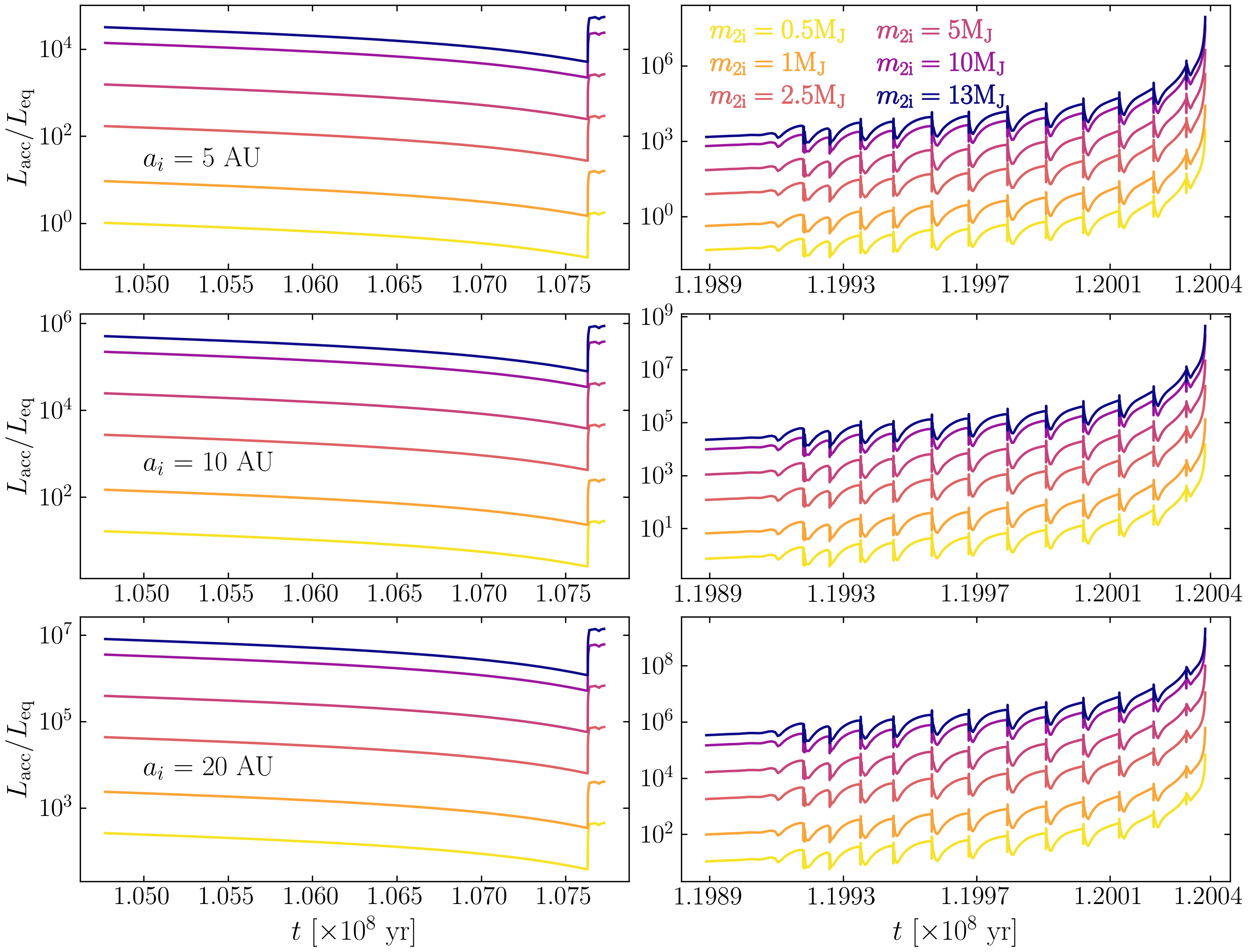}
\captionsetup{justification=justified} 
\caption{Evolution of the ratio $L_{\mathrm{acc}}$/$L_{\mathrm{eq}}$ during the RGB phase (left panels) and TPAGB phase (right panels). Each row corresponds to a different initial semi-major axis: $a_\mathrm{i}=5$, 10 and 20~AU from top to bottom, respectively.}
\label{Figprueba}
\end{figure*}

We also evaluated the ratio $L_{\mathrm{acc}}$/$L_{\mathrm{eq}}$ for the simulated planets during the RGB and TPAGB phases of the 2 $\mathrm{M}_\odot$ stellar model. The results are shown in Fig.~\ref{Figprueba}, where the left and right panels correspond to the RGB and TPAGB phases, respectively. Each row represents a different initial orbital separation (5, 10, and 20 AU), and different colors indicate the planetary masses. With the exception of the model with $a_i=5$~AU and $m_{2,i}=0.5$ M$_\mathrm{J}$, all simulated planets satisfy $L_{\mathrm{acc}}$/$L_{\mathrm{eq}} > 1$, with higher ratios during the TPAGB phase. The ratios reach up to $\sim 10^{9}$, confirming that the accretion luminosity overwhelmingly dominates over the equilibrium component and may, in principle, be observable in real planetary systems. These results are qualitatively consistent with the analytical estimates of \citet{SPMAD}, who showed that the accretion luminosity of giant planets embedded in RGB and AGB winds can exceed the luminosity intercepted from stellar irradiation. Although the detectability of planets embedded in RGB or AGB winds is strongly limited by the large contrast with the host star and by circumstellar dust emission, it is instructive to compare our predictions with existing observational constraints. Nevetherless, there are other observational signatures caused by planetary engulfment or tidal destruction events,  that may lead to the detection of a planet in these type of systems \citep{2018Galax...6...58S, 2023Natur.617...55D}

To date, several exoplanets orbiting MS stars have had their thermal emission directly detected, including systems such as HR~8799~b–e \citep{Marois2010}, $\beta$~Pictoris~b \citep{Lagrange2010}, 51~Eridani~b \citep{Macintosh2015}, and GJ~504~b \citep{Kuzuhara2013}. Additional flux detections have been obtained for hot Jupiters through infrared phase-curve and secondary-eclipse measurements with the {\it Spitzer Space Telescope}, such as HD~209458~b \citep{Deming2005} and HD~189733~b \citep{Knutson2007}. These planets exhibit effective temperatures of several hundred to several thousand kelvin and orbit relatively bright MS hosts, conditions that differ substantially from the much cooler and more extended environments characteristic of RGB and AGB stars

More recently, the {\it James Webb Space Telescope (JWST)} enabled the first direct detection of thermal emission from a planet orbiting a WD. \citet{Limbach2025} reported emission from WD~1856+534~b, with a measured temperature of approximately 186~K and a luminosity of $\sim10^{-9}\, \mathrm{L}_\odot$. This represents the coldest and faintest exoplanet ever detected in emission, establishing the current lower limit in planetary luminosity accessible to infrared observations. The larger values of the accretion luminosities predicted by our simulations, ranging from $10^{22}$ to $10^{31}$~erg~s$^{-1}$ ($10^{-11}$–$10^{-3}\,\mathrm{L}_\odot$), therefore exceed by several orders of magnitude the faintest planetary flux now measurable. Although the high luminosities and dusty envelopes of RGB and AGB stars pose severe contrast challenges, the sensitivity demonstrated by JWST indicates that accretion-powered emission at the levels predicted here could, in principle, be detectable with current or forthcoming infrared facilities. This comparison highlights that, while direct flux measurements of planets around luminous giants remain observationally demanding, they are no longer beyond the reach of present-day instrumentation.

\subsection{Dynamical evolution with and without stellar wind accretion and stellar wind drag force}

To assess the dynamical influence of stellar wind accretion and the associated stellar wind drag force on planetary orbits, we performed a set of control simulations aimed at isolating these effects. To quantify this, we computed the difference in the final orbital distance for the system that accreted the most mass ($a_\mathrm{i}$ = 5 AU and $m_\mathrm{2i}$ = 13 $\mathrm{M}_\mathrm{J}$), with and without wind accretion, and found a variation of only $\Delta a_{\mathrm{f}}$ = 1.98 $\times 10^{-3}$ AU. This confirms that the orbital expansion is primarily driven by stellar mass loss rather than by the small amount of mass accreted by the planet, result that was implied in calculations from \citet{1998Icar..134..303D}.

We also examined the potential role of the stellar wind drag force in modifying the orbital evolution. For this test, we explicitly include a drag term of the form

\begin{equation}
\vec{F}_\mathrm{drag}=\dot{M}_\mathrm{acc}~\vec{v}_\mathrm{rel}, 
\end{equation}
where $v_\mathrm{rel}$ is the relative velocity between the stellar wind and the velocity of the planet \citep[see][]{TejedaToala2025,Maldonado2025b}. The test was performed on the same system ($m_{\mathrm{2i}} = 13~\mathrm{M}_\mathrm{J}$ and $a_{\mathrm{i}} = 5$~AU), incorporating both mass accretion and stellar wind drag force. The results show that including this drag term has a negligible effect on the orbital evolution, i.e.~the planet's orbital distance remains unchanged relative to the simulation including accretion alone.

Together, these tests demonstrated that the evolution of the planetary orbits in our models is governed almost entirely by stellar mass loss. In the most extreme accretion case, the inclusion of the stellar wind drag force produced no measurable impact on the orbital evolution. Therefore, this test was not extended to systems with less massive planets or wider orbital separations, where the effect of the drag force would be even smaller. Both stellar wind accretion and the corresponding wind drag force thus contribute negligibly to the overall dynamical behavior and can be safely neglected in subsequent analyses.

These results are consistent with previous studies showing that, in the planetary-mass regime, orbital evolution during post-main-sequence mass loss is dominated by the decrease in stellar mass rather than by mass transfer \citep[e.g.,][]{Villaver2009}. Conversely, for compact or more massive companions, such as WDs, the drag force can play a much more significant role, even reversing the outward migration caused by stellar mass loss, as recently demonstrated by \citet{Maldonado2025b}. Our findings therefore reinforce the view that, for planetary companions, wind accretion and drag remain dynamically unimportant compared to the dominant influence of stellar mass loss.

\subsection{Stellar wind accretion in planetary systems with two planets} \label{sec4.6}

To further investigate stellar wind accretion in observed planetary systems orbiting evolved Solar-like stars, we extended our analysis to systems hosting two confirmed exoplanets. For each of these systems, we performed additional simulations over ten orbital periods of the outermost planet. The stellar and planetary parameters were compiled from \citet{2023Chen}, the NASA Exoplanet Archive \citep{2013akeson} and the Extrasolar Planets Encyclopaedia \citep{Schneider2011}. The corresponding accretion and equilibrium luminosities, along with their ratios {$L_{\mathrm{acc}}$/$L_{\mathrm{eq}}$}, are listed in Table~\ref{tab:tab3}. 

\begin{table}
\centering
\resizebox{\columnwidth}{!}{%
\begin{tabular}{|l|l|l|l|l|l|l|l|}
\hline
 \multirow{2}{*}{Exoplanet} & $m_{\mathrm{2}}$ & $a$ & $L_{{\mathrm{acc}}}$ &  $L_{{\mathrm{eq}}}$ & \multirow{2}{*}{$L_{\mathrm{acc}}$/$L_{\mathrm{eq}}$}\\
 & ($\mathrm{M}_\mathrm{J}$)  & (UA) & (erg s$^{-1}$) & (erg s$^{-1}$) & \\
\hline
HD 75784 b  &   1.15  &  1.032  &   $1.31\times10^{26}$ &  $2.96\times10^{21}$ & $4.42\times10^{4}$ \\
HD 75784 c  &   5.64  &  8.4    &   $1.25 \times10^{28}$ &  $5.60\times10^{17}$ & $2.23\times10^{10}$ \\ 
\hline
HD 1605 b &   0.934 &  1.48   &   $6.14\times10^{25}$ &  $1.35 \times10^{21}$ & $4.54\times10^{4}$ \\
HD 1605 c   &   3.62  &  3.52   &   $3.55 \times10^{27}$ &  $3.35 \times10^{19}$ & $1.05\times10^{8}$ \\
\hline
\end{tabular}
}
\captionsetup{justification=justified} 
\caption{Planetary parameters, along with the predicted accretion and equilibrium luminosities after 10 orbits of the outermost planet for each of the two planetary systems discussed in Section~\ref{sec4.6}, the ratio of this luminosities is presented as well. The stellar parameters are presented in Table \ref{tab:2}.}
\label{tab:tab3}
\end{table}

For these calculations, we adopted a constant stellar mass-loss rate of $\dot{M}_\mathrm{w} = 5 \times 10^{-7}$~$\mathrm{M}_\odot$~yr$^{-1}$, consistent with the stellar evolution model for a  2~$\mathrm{M}_\odot$ star during the RGB phase (see Fig. \ref{Fig:2}). The results indicate that, even in multi-planet configurations and under the current system parameters, the accretion luminosity of each planet exceeds its equilibrium luminosity after ten orbits of the outermost companion, which also corresponds to the most massive planet in each systems. Interestingly, despite being the outermost companions, these planets exhibit the highest accretion rates, consistent with their larger masses. These findings suggest that, for the present orbital architectures of such systems, accretion luminosity could be observationally detectable.

Although these two systems provide valuable insight into the potential observability of stellar wind accretion in two-planet systems, their small number and short integration timescales, limited to ten orbital periods, restrict the scope of our conclusions. The long-term dynamical response of two-planet systems to stellar mass loss has been investigated in several studies (e.g., \citealt{Veras2013}; \citealt{2013V}), which showed that mass loss can induce instabilities when planets lie near the Hill or Lagrange stability boundaries, leading to scattering or orbital rearrangements. Building upon these theoretical efforts, \citet{Maldonado2020} analyzed scaled analogues of observed two-planet systems and found that only a few developed dynamical instabilities during the WD phase. This indicates that most realistic systems remain dynamically stable throughout the RGB and AGB evolution. Our results, which show short-term stability and enhanced accretion in the outermost massive planet, are consistent with this broader picture. Extending these models to longer timescales and a wider range of planetary architectures, including self-consistent wind accretion and drag, will be essential to determine whether this behavior is generic or system-dependent.

\subsection{Caveats}
Exoplanets orbiting evolved Solar-like stars can exhibit eccentric orbits ($e > 0$; see Table \ref{tab:2}). As a first step toward modeling stellar wind accretion in planetary systems with evolving Solar-like host stars, we have restricted our analysis to circular orbits ($e = 0$). Eccentric orbits require a more detailed treatment since reduced orbital separation near and at periastron may activate different accretion regimes and enhance dynamical effects, such as tidal interactions or WRLOF, which are beyond the scope of this study. We plan to address eccentric configurations in future work, using the framework developed here as a baseline, and the mass accretion efficiency formalisms presented for eccentric orbits in \citet{TejedaToala2025}.

\section{SUMMARY AND CONCLUDING REMARKS} \label{sec5}
\label{sec:summary}

Wind accretion onto planets orbiting evolving Solar-type stars remained relatively unexplored, even though similar processes have been extensively studied in the context of stellar binaries. The velocity regime where the wind and orbital velocities are comparable ($v_\mathrm{w}\lesssim v_\mathrm{o}$) has not yet been investigated in detail for planetary companions, even though such conditions are expected to occur in systems around evolved stars. In this work, we applied the modified BHL implementation proposed by \citet{TejedaToala2025}, which provides an improved description of wind accretion in this regime, to study the accretion behavior of planetary systems during the RGB and TPAGB phases.

We performed a suite of simulations to explore the evolution of planetary systems around an evolving 2~$\mathrm{M}_{\odot}$ star during its giant phases. The sample includes 18 planetary system configurations spanning planetary masses between 0.5 and 13~$\mathrm{M}_\mathrm{J}$ and initial orbital separations of 5, 10, and 20~AU. Stellar evolution tracks were computed with MESA, and the  orbital dynamics were integrated with REBOUND. This framework enabled a consistent treatment of stellar mass loss, wind properties, and the resulting planetary orbital evolution and accretion behavior. 

Our conclusions can be summarized as follows: 
\begin{itemize}
    \item The mass accretion rate of a planet increases with decreasing semi-major axis and increasing planetary mass. Throughout most of the evolution, $w =v_\mathrm{w}/v_\mathrm{o}<1$, and the accretion efficiency $\eta$ remains nearly constant. As the systems enter the TPAGB phase, stellar mass loss and orbital expansion increase $w$, while the accretor's gradual mass growth raises the mass ratio $q$. The interplay between these effects leads to a mild overall increase in $\eta$ during the final stages of evolution. 
    \item In the most compact planetary systems ($a_\mathrm{i} = 5$ AU), the total mass accreted by each planet is less than 0.6 $\%$ of its initial mass. This small accreted fraction is insufficient to counteract the orbital expansion caused by stellar mass loss. Since this corresponds to the most favorable case for mass accretion, the same conclusion applies to all other configurations with lower planetary masses or wider separations. The dynamical evolution of the systems is therefore governed primarily by the mass loss of the host star.  
    \item The estimated equilibrium luminosities of the 21 exoplanets listed in Table~\ref{tab:2} fall within the range of our simulation results, indicating that the properties of the simulated planetary systems are broadly consistent with those inferred for observed exoplanetary systems.
    \item The accretion luminosity exceeds the equilibrium luminosity in the majority of the simulated cases. This suggests that, in principle, the accretion luminosity of real exoplanets could be observable during the RGB and TPAGB phases of Solar-like stars, with any potential planetary emission being dominated by accretion processes.

    \item Tests including the stellar wind drag force show that, within the explored parameter range, its effect on the dynamical evolution of the planet is negligible. In the few simulated two-planet configurations, the more massive companions accrete larger amounts of stellar wind material than their lower-mass counterparts, following trends consistent with those found in single-planet systems. While these results support the overall robustness of our conclusions, a more detailed study covering a broader range of planetary architectures will be needed to confirm these behaviors.  
    
\end{itemize}

\renewcommand{\refname}{REFERENCES}

\bibliography{rmaa_2}

\section{ACKNOWLEDGEMENTS}

PPL, JAT and JBR-G. acknowledge support from the UNAM PAPIIT project IN102324. RFM thanks UNAM DGAPA (Mexico) and SECIHTI (Mexico) for postdoc fellowships. JAT thanks the staff of Facultad de Ciencias de la Tierra y el Espacio of Universidad Autónoma de Sinaloa (FACITE-UAS, Mexico) for their support during a research visit. This work has made extensive use of NASA's Astrophysics Data System.

\end{document}